\documentclass[12pt,a4paper]{article}
\usepackage{amsmath}
\usepackage{amssymb}
\usepackage{graphicx}
\usepackage{url}
\usepackage{wrapfig}
\usepackage{multicol}
\usepackage{multirow}
\usepackage{authblk}
\usepackage{esint}
\usepackage{floatflt}
\usepackage[font=small,labelfont=bf]{caption}
\usepackage{subcaption}
\usepackage{mathtools}
\usepackage{sectsty}
\usepackage{hyperref}
\usepackage{appendix}
\usepackage{cite}
\hypersetup{
    colorlinks=true,
    allcolors=blue,
}
\newcommand{\eq}[2]{\begin{equation} \label{#1} #2 \end{equation}}
\newcommand{\meq}[2]{\begin{equation}\begin{aligned} \label{#1} #2 \end{aligned}\end{equation}}

\title{\bf Surface reaction-diffusion kinetics on lattice at the microscopic scale}
\date{}

\author{\small Wei-Xiang Chew$^{1,2}$, Kazunari Kaizu$^{1}$, Masaki Watabe$^{1}$, Sithi V. Muniandy$^{2}$, Koichi Takahashi$^{1}$ and Satya N. V. Arjunan$^{1*}$}
\affil{\small
$^{1}$ Laboratory for Biologically Inspired Computing, RIKEN Center for Biosystems Dynamics Research, Suita, Osaka, Japan\\
$^{2}$ Department of Physics, Faculty of Science, University of Malaya, 50603, Kuala Lumpur, Malaysia.\\
$^{*}$E-mail: satya@riken.jp 
}

\begin{document}
\maketitle
\sectionfont{\large\MakeUppercase}

\section*{Abstract}
Microscopic models of reaction-diffusion processes on the cell membrane can link local spatiotemporal effects to macroscopic self-organized patterns often observed on the membrane. Simulation schemes based on the microscopic lattice method (MLM) can model these processes at the microscopic scale by tracking individual molecules, represented as hard-spheres, on fine lattice voxels. Although MLM is simple to implement and is generally less computationally demanding than off-lattice approaches, its accuracy and consistency in modeling surface reactions have not been fully verifed. Using the Spatiocyte scheme, we study the accuracy of MLM in diffusion-influenced surface reactions. We derive the lattice-based bimolecular association rates for two-dimensional surface-surface reaction and one-dimensional volume-surface adsorption according to the Smoluchowski-Collins-Kimball model and random walk theory. We match the time-dependent rates on lattice with off-lattice counterparts to obtain the correct expressions for MLM parameters in terms of physical constants. The expressions indicate that the voxel size needs to be at least $0.6\%$ larger than the molecule to accurately simulate surface reactions on triangular lattice. On square lattice, the minimum voxel size should be even larger, at $5\%$. We also demonstrate the ability of MLM-based schemes such as Spatiocyte to simulate a reaction-diffusion model that involves all dimensions: three-dimensional diffusion in the cytoplasm, two-dimensional diffusion on the cell membrane and one-dimensional cytoplasm-membrane adsorption. With the model, we examine the contribution of the 2D reaction pathway to the overall reaction rate at different reactant diffusivity, reactivity and concentrations.
\newpage

\section{Introduction}
\paragraph{}

Many essential and intriguing intracellular biochemical systems are mediated by the cell membrane. These systems include cell polarity establishment \cite{Halatek2018}, symmetrical cell division \cite{Wettmann2018}, modulation of signal transduction \cite{Zhou2018} and directed cell migration \cite{Devreotes2017}. Spatiotemporal patterns arising from protein self-organization on the membrane \cite{Kretschmer2016}, play a central role in these systems. The proteins self-organize primarily by reaction and diffusion processes. Membrane interactions can be classified as surface-surface reactions, where membrane-bound molecules react with each other; and volume-surface reactions, where cytosolic molecules react with membrane lipids or membrane-bound molecules. 

To uncover the mechanisms underlying these systems, reaction-diffusion modeling approaches have been useful \cite{Pablo2018,Denk2018,Krishnan2012,Tian2010,Arai2010}. In general, the choice of modeling approach depends on the time and length scales of the system \cite{Takahashi2005, Berro2018}. When the molecule copies are abundant and are well mixed in the surface compartment, macroscopic modeling approaches that apply rate \cite{edelstein1988mathematical} or reaction-diffusion \cite{edelstein1988mathematical,murray2001mathematical} equation, are sufficient. If the molecule number is small or when the molecules are not homogeneously distributed in the compartment, mesoscopic methods based on the reaction-diffusion master equation (RDME) \cite{gillespie2014validity,isaacson2018unstructured,hellander2017reaction,smith2018spatial} can be employed since they account for both the fluctuations from small number of molecules and the spatial inhomogeneity across well mixed compartment subvolumes \cite{fange2006noise,lawson2013spatial}. 

Although macroscopic and mesoscopic approaches are applicable for large scale simulations, the well mixed assumption imposes several limitations. These approaches for example, cannot explicitly capture the effects of space at the microscopic scale that arise from the interactions of finite-sized molecules \cite{almeida1995lateral,melo2006kinetics,Guigas2016}, fast rebinding of reactants \cite{lagerholm1998theory,loose2011min,mugler2012membrane} and microscopic surface inhomogeneity such as lipid domains and membrane-associated cytoskeletal structures \cite{almeida1995lateral,kalay2012confining,kalay2012reaction,zhdanov2015kinetics,kerketta2016effect}. The spatial effects can alter not only the diffusion behavior \cite{nicolau2007sources,dix2008crowding,hofling2013anomalous}, but also the reaction kinetics \cite{saxton1997single,melo2006kinetics,kalay2012reaction,zhdanov2015kinetics,kerketta2016effect}, leading to different physiological outcomes. For example, clustering of membrane receptors changes the response of signaling network \cite{mugler2012membrane}, fluctuation in protein copy number promotes cell polarization in the absence of spatial cue \cite{Pablo2018}, rapid protein rebinding affects spatiotemporal patterns on the membrane \cite{loose2011min} and amplifies noise during ligand interactions \cite{kaizu2014berg}. Moreover, macroscopic and mesoscopic approaches adopt the macroscopic reaction rate constant for all reactions, which is not sufficient for irreversible bimolecular surface-surface and one-dimensional volume-surface reactions because these time-dependent reaction rates do not reach steady state \cite{szabo1989,Yogurtcu2015}. 

Microscopic approaches are more suitable to model surface interactions for short timescales when the microscopic spatial effects need to be directly accounted for \cite{berry2002monte,saxton2007modeling,schoneberg2014simulation,Trovato2017,Andrews2018}. In off-lattice microscopic particle-based methods, diffusion is simulated in continuous space with Gaussian distributed displacement. Bimolecular reactions are executed according to the Smoluchowski-Collins-Kimball (SCK) \cite{smoluchowski1917,Collins1949} or the Doi \cite{doi1976stochastic,erban2009stochastic,agbanusi2014comparison} physical model. In the former, the reaction occurs either immediately or with a probability of reflection when the distance between reactants equals to a predefined reaction radius, whereas in the Doi model, the reaction occurs with a fixed probability per unit time when the reactants are closer than the radius. Off-lattice SCK methods that support surface reactions include Smoldyn \cite{Andrews2009,andrews2016smoldyn}, CDS \cite{Byrne2010} and eGFRD \cite{sokolowski2017egfrd}, while the Doi model is adopted by MCell \cite{Kerr2008}, ReaDDy \cite{schoneberg2013,Hoffmann374942} and SpringSaLaD \cite{michalski2016}. Smoldyn also recently included the option to support the Doi model. All of these methods except MCell can simulate volume occupying molecules. In a recent performance benchmark that did not include CDS \cite{Andrews2018}, Smoldyn displayed the fastest simulation run time for a simple enzymatic reaction model in volume compartment. 

Schemes based on the microscopic lattice method\footnote{also called Monte Carlo lattice gas model \cite{saxton2007modeling}} (MLM) \cite{saxton2007modeling,chew2018} attempt to reduce the cost of resolving molecular collisions by discretizing the space into fine molecule-sized voxels. In the Spatiocyte scheme for example, a molecule only checks its destination voxel for occupancy before performing a bimolecular reaction with the occupying molecule or moving into it if it is vacant \cite{Arjunan2010,chew2018}. Consequently, Spatiocyte exhibits better run time and scaling performances than Smoldyn when diffusing volume occupying molecules \cite{chew2018}. The run time of Spatiocyte is also comparable to Smoldyn in the benchmark enzymatic reaction model \cite{chew2018}. The reduced computational cost and the simplicity of MLM implementation have promoted its applications in both biological \cite{Berry2002a,Schnell2004,saxton2008biological,Arjunan2010,tsourkas2011monte,Shimo2015} and chemical \cite{lombardo1991review,imbihl1995oscillatory,lukkien1998efficient} surface reactions. Nonetheless, biological surfaces such the cell and nuclear membranes are not arranged as fixed lattice structures. Further, since diffusion and reaction kinetics can be influenced by the lattice arrangement \cite{hughes1995random,saxton1987lateral,meinecke2016}, the accuracy of MLM compared to off-lattice particle-based methods requires careful examination. Notably, a consistent approach is needed to determine MLM parameters such as voxel size and on-lattice reaction probability that can replicate the kinetics in continuous space.

In previous work, the SCK model was used together with the Spatiocyte scheme to construct a general theoretical framework of MLM for simulating reaction and diffusion processes in three-dimensional (3D) space \cite{chew2018}. Within this framework, the expressions for on-lattice reaction rate constant, reaction and rebinding probabilities, and voxel size were derived to reproduce off-lattice reaction kinetics consistently. Here, we extend this framework for two-dimensional (2D) surface-surface and one-dimensional (1D) volume-surface reactions. We also employ the SCK model to derive on-lattice time-dependent rate coefficients for the surface reactions. We then obtain the expressions for the MLM parameters by equating the off-lattice rate equations with the on-lattice counterparts. In section \ref{s1} of this paper, we introduce the existing continuum-based reaction kinetics theory for surface reactions. In section \ref{s2}, we derive the expressions for surface reaction rates on lattice according to the Spatiocyte scheme and verify them using the continuum theory. In section \ref{s3}, we demonstrate the applicability of the derived expressions for surface reactions that involve all dimensions. We also look at the contribution of 3D and 2D reaction pathways to the overall reaction rate. Finally in section \ref{s4}, we discuss the implications and limitations of this work. 

\section{Continuum-based reaction kinetics theory} \label{s1}
\paragraph{}
Consider a many-body bimolecular reaction,
\eq{irrrxn}{A+B \xrightarrow{} B,}
with $A$ and $B$ having radii $r_A$ and $r_B$, and diffusion coefficients $D_A$ and $D_B$, respectively. According to the SCK model, when the distance between a pair of $A$ and $B$ molecules is the sum of their radii $R=r_A+r_B$, the two will react with an intrinsic rate $k_a$. The fraction of $A$ remaining in the system is described by the survival probability, $S_{irr,A}(t)=[A(t)]/[A(0)]$, where [ ] denotes the concentration. When $[B(0)]\gg[A(0)]$, the survival probability of $A$ is provided in the rate equation \cite{szabo1989}:
\eq{STirrA}{\frac{dS_{irr,A}(t)}{dt}=-[B]k(t)S_{irr,A}(t),} 
where $k(t)$ represents the time-dependent rate coefficient. The solution for the survival probability requires the integration of the rate coefficient (Eq. 2.35 in \cite{szabo1989}):
\eq{stirr}{S_{irr,A}(t,[B])=\exp\left[-[B]\int_0^t k(\tau)d\tau \right].}

According to the particle-pair formalism of SCK model \cite{Noyes1954,berg1978,naqvi1980,tachiya1983theory}, the many-body reaction can be approximated by a simpler two-body problem. The time-dependent rate coefficient can thus be expressed as the product of $k_a$ and the survival probability of an in-contact reactant pair: 
\eq{ktnoyes}{k(t) = k_a\left[1- \int_0^t  p_{reb}(R,\tau|R,0)d\tau\right].}  
Here, $p_{reb}(R,\tau|R,0)$ specifies the rebinding-time probability distribution for a reactive particle-pair separated by a distance $R$ at time $\tau$, given that the pair were initially in contact. 

The specific functional form of the rate coefficient depends on the spatial dimension of the reactant diffusion. The dimension is 2D for surface-surface reactions, whereas for volume-surface reactions, it is determined by the target reactant of the cytosolic molecule. The dimension is 3D when the reactant is a membrane-associated molecule or 1D when the cytosolic molecule reacts directly with the lipid membrane. For clarity, we refer to the 1D volume-surface reaction as adsorption. Since we have previously described the theory for 3D reactions \cite{chew2018}, in the following subsections we provide the theory for 2D surface-surface reaction and 1D volume-surface adsorption.

\subsection{2D surface-surface reaction}
\subsubsection{Irreversible reaction}
\paragraph{}
The time-dependent rate coefficient for 2D association reaction with radiation boundary condition is given in the Laplace form as \cite{szabo1989}:
\eq{k2ds}{
k_{2D}(s)=\frac{k_{a2D}}{s[1+k_{a2D}\widetilde{g}(s)]}.
}
Here, $k_{a2D}$ is the intrinsic rate constant with dimensions of length, L and time, T, given by $\mathrm{L^2T^{-1}}$, and $\widetilde{g}(s)$ is the Green's function for 2D diffusion defined as \cite{popov2002exact}
\eq{2DGF}{\widetilde{g}(s)=\frac{K_0(\tilde{s})}{2\pi D \tilde{s} K_1(\tilde{s})}.
}
$K_0$ and $K_1$ are the modified Bessel functions of the second kind, $\tilde{s}=\sqrt{sR^2/D}$, and $D=D_A+D_B$. 

Eq. \eqref{k2ds} can thus be written as
\eq{k2ds1l}{k_{2D}(s)=2\pi D\tilde{s}\frac{K_1(\tilde{s})}{s[K_0(\tilde{s})+2\pi\tilde{s}K_1(\tilde{s})/\kappa]},}
with $\kappa=k_a/D$. In the limit of small $\tilde{s}$, we can approximate the modified Bessel functions:
\eq{}{\tilde{s}K_1(\tilde{s})\approx 1-\frac{2\ln(\tilde{s}e^\gamma /2)+1}{4}\tilde{s}^2+O(\tilde{s}^4),}
and
\eq{}{K_0(\tilde{s})\approx -\ln(\tilde{s}e^\gamma /2)-\frac{2\ln(\tilde{s}e^\gamma /2)+1}{4}\tilde{s}^2+O(\tilde{s}^4),}
where $\gamma=0.5772156$ is the Euler constant.

Using these approximations, the asymptotic expansion of Eq. \eqref{k2ds1l} can be expressed as:
\meq{asympks}{
\lim_{s\to 0}sk_{2D}(s)&=\frac{4\pi D}{-2\ln{[R\sqrt{s/D}]}+\ln{4}+\ln{[\exp{2(4\pi /\kappa-\gamma)}]}}+O(s)\\
&=\frac{4\pi D}{\ln{[C_c/s]}}+O(s)\\
}
with $C_c=4D\exp(4\pi /\kappa-2\gamma)/R^2$. The corresponding long-time approximation is given as \cite{ritchie1956}:
\meq{kck}{k_{2D}(t\gg\frac{R^2}{D})&=4\pi D(\frac{1}{\ln(C_ct)}-\frac{\gamma}{(\ln(C_ct))^2}-\frac{1.311}{(\ln(C_ct))^3}\\
&+\frac{0.25}{(\ln(C_ct))^4}+O(t^{-1}\ln(t)^{-2}))
,}
where the relative error to the exact form is less than $1\%$ at $t=100R^2/D$. Note that unlike in 3D space \cite{chew2018}, the long-time approximation of the 2D rate coefficient does not converge to a constant term as $t\to \infty$. 

\subsubsection{Steady-state rate constant}
\paragraph{}
It is convenient to describe the 2D association reaction with a single characteristic rate constant. This is possible by defining a steady-state rate constant in terms of the mean lifetime of $A$, $\tau_m$ \cite{szabo1989}: 
\meq{}{
k_{ss}&=\frac{1}{[B]\tau_m}\\
&=\left[[B]\int_0^\infty S_{irr,A}(t) dt \right]^{-1}\\
&=\left[[B]\int_0^\infty \exp\left(-[B] \int_0^t k_{2D}(\tau)d\tau \right)dt \right]^{-1}.\\
}
$k_{ss}$ can be evaluated using the mean field self-consistency condition \cite{szabo1989}:
\eq{mfkss}{k_{ss}=[s\hat{k}(s)]_{s=k_{ss}[B]}.} Substituting the asymptotic form of $k_{2D}(s)$, as defined in Eq. \eqref{asympks}, into Eq. \eqref{mfkss} yields
\eq{}{\frac{k_{ss}}{2\pi D}\approx \left[\ln2 -\gamma-\ln[R\sqrt{k_{ss}[B]/D}]+1/\kappa \right]^{-1}.} Rewriting some variables in terms of the molecule area fraction, $\phi=\pi R^2[B]$ and taking the small concentration limit, $\phi \to 0$ gives the following approximation
\meq{}{\frac{k_{ss}}{2\pi D}&\approx \left[\ln2 -\gamma-\ln[\sqrt{2\phi y}]+1/\kappa \right]^{-1}\\
&\approx \left[\frac{1}{2}\ln2-\gamma-\frac{1}{2}\ln\phi+1/\kappa \right]^{-1}.\\
}
Finally, the steady-state rate constant for radiation boundary condition is obtained as
\eq{kss2d}{
k_{ss2D}\approx \frac{4\pi D}{ \ln2-2\gamma-\ln\phi+4\pi /\kappa}.} 
Similar to the 3D effective rate constant in $1/k_{ss3D}=1/k_a+1/(4\pi RD)$, the 2D steady-state rate constant depends on the intrinsic rate $k_a$ and the relative diffusion coefficient $D$. Interestingly, the 2D rate constant has the additional dependency on the concentration of $B$.

\subsubsection{Reversible reaction}
\paragraph{}
In the SCK model for 2D reversible reaction 
\eq{rerxn}{A+B \underset{k_{d2D}}{\stackrel{k_{a2D}}{\rightleftharpoons}} AB,}
a bound pair $A$-$B$ dissociates with the rate constant $k_{d2D}$ ($\mathrm{T^{-1}}$) into $A$ and $B$, separated at distance $R$. The survival probability of $A$, defined as $S_{rev,A}(t)$, can be calculated using the first variant of the multiparticle kernel theory (MPK1) \cite{sung1999,popov2001}. Although the closed form solution for $S_{rev,A}(t)$ in 2D is not available, it can be evaluated by numerically solving the normalized deviation defined as 
\eq{Stre}{\frac{S_{rev,A}(t)-S_{rev,A}(\infty)}{S_{rev,A}(0)-S_{rev,A}(\infty)}=\mathcal{L}^{-1}\left[\frac{1}{s+\lambda \widetilde{F}(s)} \right].}
Here, the term
\eq{}{\widetilde{F}(s)=\frac{k_{d2D}}{\lambda}\widetilde{F}_{gem}(s)+\frac{[B]k_{a2D}}{\lambda}\widetilde{F}_{irr}(s;[B]')}
is the diffusion factor function, $S_{rev,A}(\infty)=k_{d2D}/\lambda=1/(1+[B]k_{a2D}/k_{d2D})$ is the equilibrium value, $\lambda=k_{d2D}+[B]k_{a2D}$ is the chemical kinetics relaxation rate constant, and $[B]'=\lambda/k_{a2D}$ is the modified concentration. $\widetilde{F}_{gem}(s)=1+k_{a2D}\widetilde{g}(s)$ contains the 2D Green's function term $\widetilde{g}(s)$ as given in Eq. \eqref{2DGF}, whereas the function
\eq{}{\widetilde{F}_{irr}(s;c_0)=\frac{c_0k_{a2D}\widetilde{S}_{irr,A}(s;c_0)}{1-s\widetilde{S}_{irr,A}(s;c_0)}, } uses the term $\widetilde{S}_{irr,A}(s;c_0)$, which is the Laplace transform of the irreversible reaction survival probability, $S_{irr,A}(t;c_0)$.

\subsection{1D volume-surface adsorption}
\paragraph{}
Before describing the rate for volume-surface adsorption, we first consider the 1D SCK model, where a single immobile $B$ interacts with multiple mobile $A$ on a filament according to Eq. \eqref{irrrxn}. $A$ can collide with $B$ from both sides of $B$, while there is no self interaction among $A$ molecules. The time-dependent rate coefficient of this reaction with radiation boundary condition is given as \cite{szabo1989}
\eq{1dradt}{k_{1D}(t)=k_{sa}\exp(\kappa^2 Dt/4)\text{erfc}(\kappa\sqrt{Dt/4}),} 
with $\kappa=k_a/D$ denoting the ratio between the intrinsic adsorption rate constant $k_a=k_{sa}$ (unit $\mathrm{LT^{-1}}$) and the relative diffusion coefficient  $D$. 
At long time, Eq. \eqref{1dradt} behaves asymptotically as
\eq{kt1dlt}{k_{1D}(t\to \infty)\approx 2\sqrt{\frac{D}{\pi t}}+O(t^{-3/2}).}

Next we consider a volume-surface adsorption system that involves an adsorbing plane at $x=0$ and bulk molecules at $x>0$. Initially, the molecules of concentration $c_0$ are distributed uniformly in the bulk and are absent on the surface. For surface adsorption process that follows the radiation boundary condition, the number of molecules adsorbed to the surface varies as (Eq. 3.37 of \cite{crank1979}):
\eq{Nsa}{N_s(t)=\frac{c_0S}{\kappa}\left\{\exp (\kappa^2Dt)\text{erfc}(\kappa\sqrt{Dt})-1+2\kappa\sqrt{Dt/\pi} \right\},}
where $S$ is the area of the plane.
The corresponding adsorption rate is well-described by the time-dependent adsorption rate coefficient $k_{sa}(t')$:
\meq{adbrate}{\frac{dN_s(t')}{dt}&=k_{sa}(t')c_0S.
}
Note that the adsorption rate coefficient differs from the 1D SCK rate by a factor of two: $k_{sa}(t')=k_{1D}(t)/2$, because in the latter, it occurs on both sides of the plane.
At long time, the adsorption rate coefficient becomes
\eq{ktadlt}{k_{sa}(t\to \infty)\approx \sqrt{\frac{D}{\pi t}}+O(t^{-3/2}).}

As the bulk molecules are adsorbed to the surface, a spatial concentration gradient develops in the bulk. The spatialtemporal concentration profile of the bulk molecules $C(x,t)$ follows Eq. 3.35 of \cite{crank1979}:
\eq{gradient}{C(x,t)=c_0\left[\text{erfc}\frac{x}{2\sqrt{Dt}}-\exp \left(\frac{k_{sa}x}{D}+\frac{k_{sa}^2t}{D}\right)\text{erfc}\left(\frac{x}{2\sqrt{Dt}}+k_{sa}\sqrt{\frac{t}{D}} \right) \right].}
When adsorbed molecules can dissociate from the surface with a rate $k_{sd}$ $(\mathrm{T^{-1}})$, their number varies according to (Eq. A.12 in \cite{Andrews2009})  
\eq{Nsad}{N_s(t)=\frac{c_0Sk_{sa}t\left[c_1\exp(c_2^2)\text{erfc}(c_2)-c_1+c_2\exp(c_1^2)\text{erfc}(c_1)-c_2 \right]}{c_1c_2(c_2-c_1)},}
where 
\eq{}{c_1=\frac{k_{sa}-\sqrt{k_{sa}^2-4Dk_{sd}}}{2\sqrt{D}}\sqrt{t},}
and
\eq{}{c_2=\frac{k_{sa}+\sqrt{k_{sa}^2-4Dk_{sd}}}{2\sqrt{D}}\sqrt{t}.}

\section{Microscopic lattice method}\label{s2}
\paragraph{}
In this section, we derive the on-lattice rate coefficients for 2D surface-surface reaction and 1D volume-surface adsorption based on the Spatiocyte scheme \cite{Arjunan2010,chew2018}. The particle-pair formalism of SCK model and random walk theory will be used in the derivation. In MLM, the SCK rate coefficient in Eq. \eqref{ktnoyes} is discretized into (see Section II.C of \cite{chew2018} for derivation): 
\eq{ktlattice}{k_m = k_a'\left[1-\sum_{n=0}^m H_n(s_0|s_1)\right],\ \text{for }m,n\in \mathbb{N}.}
$k_a'$ is the initial lattice rate constant (see Appendix \ref{LIR}), $H_n(s_0|s_1)$ is the rebinding-time probability distribution in diffusion step $n$, and $m$ is the simulation step. Rebinding here refers to the next reaction event of an in-contact reactant pair on lattice, $s_0$ denotes the voxel at the origin and $s_1$ refers to an element from the set of immediate neighbor voxels of $s_0$. The rebinding-time probability is a function of random walk quantities such as $P_n(s_a|s_b)$, the voxel occupation probability from voxel $s_b$ to voxel $s_a$, that is, the probability of being at voxel $s_a$ after $n$ steps, given that the walk started at voxel $s_b$; and $F_n(s_a|s_b)$, the first-passage time distribution from voxel $s_b$ to $s_a$, that is the probability of arriving at $s_b$ for the first time on the $n$th step, given that the walk started at voxel $s_a$. These quantities depend on the lattice arrangement, dimension of diffusion and also the simulation scheme. The simulation step $m$ is related to the simulation time $t'$ through the relation $2dDt'=ml^2$, where $d$ is the dimension of diffusion. In the following subsections of 2D and 1D reactions, we will derive the on-lattice rate coefficient based on the simulation scheme. 

\subsection{2D surface-surface reaction}\label{ss1}
\subsubsection{Irreversible reaction}
\paragraph{}
The methods presented in this work is generalized for any regular lattice arrangement, but we focus on the triangular lattice since it is used to simulate surface-surface reactions in the Spatiocyte scheme. The derivation of the rate coefficients for activation-limited ($k_{a2D}\ll D$) and diffusion-influenced ($k_{a2D}\gg D$) reactions is treated separately because the simulation scheme executes these two types of reaction in distinct manner. 

In the activation-limited scheme, the generating function for the rebinding-time probability distribution $H_n(s_0|s_1)$ is given as (see Appendix D.1 in \cite{chew2018} for derivation):
\eq{ALGF}{H(s_0|s_1;z)=\frac{P_aF(s_0|s_0;z)}{z+F(s_0|s_0;z)(P_a-1)},}
where $P_a$ denotes the reaction probability.
In terms of $P(s_0|s_0;z)$, the generating function of $P_n(s_0|s_0)$, $H(s_0|s_1;z)$ becomes (see Appendix \ref{ALHZ} for full derivation):
\eq{GF1}{H(s_0|s_1;z)=1-\frac{1}{P_aP(s_0|s_0;z)\left[1+\frac{(1-P_a)}{P_aP(s_0|s_0;z)}\right]}.}

The generating function $P(s_0|s_0;z)$ for the triangular lattice in asymptotic form is given as (see Appendix \ref{trilat} for details):
\eq{Pz1}{P(s_0|s_0;z)\approx \frac{\sqrt{3}}{2\pi}\ln[12(1-z)^{-1}]\{1+O(1-z)\}.}
By substituting Eq. \eqref{Pz1} into Eq. \eqref{GF1}, we obtain the following approximated form:
\eq{Hzasymp}{H(s_0|s_1;z)\approx -\frac{b_1}{P_a}\left\{\ln\left(\frac{E}{1-z}\right)\right\}^{-1},
}
where $E=12\exp\left\{b_1(1/P_a-1)\right\}$ and $b_1=2\pi/\sqrt{3}$.
We then apply singularity analysis (Figure VI.4 of \cite{flajolet2009analytic}) on Eq. \eqref{Hzasymp} to obtain the large $n$ behavior:
\eq{Hn}{H_n(s_0|s_1)\approx \frac{2\pi}{\sqrt{3}P_a}\frac{1}{n}\left(\frac{1}{(\ln En)^2}-\frac{2\gamma}{(\ln En)^3}+\frac{3\gamma^2-\frac{\pi^2}{2}}{(\ln En)^4}+... \right).}
With Eq. \eqref{Hn}, we evaluate the discrete sum in Eq. \eqref{ktlattice} using the Euler-Mascheroni formula together with the definition of recurrence in 2D random walk: $\sum_{n=0}^\infty H_n=1$.
The solution in high order $m$ terms is given by:
\meq{}{k^{2D}_m&=\frac{2\pi k_{a2D}'}{\sqrt{3}P_a}\left[\frac{1}{\ln Em}-\frac{\gamma}{(\ln Em)^2}+\frac{\gamma^2-\frac{\pi^2}{6}}{(\ln Em)^3}+... \right].}
Finally, we apply the definition of initial rate for the triangular lattice as given in Eq. \eqref{lir2d} and the relation $ml^2=4Dt'$ to obtain the time-dependent rate coefficient:
\eq{knl}{k_{2D}'(t)=4\pi D \left[\frac{1}{\ln C_lt}-\frac{\gamma}{(\ln C_lt)^2}+\frac{\gamma^2-\frac{\pi^2}{6}}{(\ln C_lt)^3}+... \right],}
where $C_l=48D\exp\left\{b_1(1/P_a-1)\right\}/l^2$ and $l$ is the voxel size.

In the derivation of diffusion-influenced scheme, it is convenient to work with the Laplace form of Eq. \eqref{ktlattice}:
\eq{lpkt}{\hat{k}'_{2D}(s)=k_{a2D}'[1-\hat{G}(s) ]/s,} 
Here $\hat{G}(s)$ is the Laplace form of the rebinding-time probability density on lattice, defined as (see Eq.D79 in \cite{chew2018}):
\eq{}{\hat{G}(s)=\beta_1 [s+\beta-sF_1(s_1|s_1)z-\beta_2F(s_1|s_1;z)]^{-1},}
where 
\eq{s1tos1}{F(s_1|s_1;z)=1-\frac{z^2P_1(s_0|s_1)}{P(s_0|s_0;z)-1},}
$P_1(s_0|s_1)=1/6$, $F_1(s_1|s_1)=1/3$, $P_1(s_2|s_1)=1/2$, $z=\beta_2/(s+\beta_2)$, $\beta=\beta_1+\beta_2$, $\beta_1=P_a/6t_d$ and $\beta_2=1/t_d$. $t_d=l^2/2dD_x$ here refers to the average time interval needed for a molecule with diffusion coefficient $D_x$ to hop across one voxel.
By applying the final value theorem, we get the asymptotic form for Eq. \eqref{lpkt} as
\meq{kslaplace}{\lim_{s\to 0}s\hat{k}'_{2D}(s)&=k_{a2D}'[1-\lim_{s\to 0}\hat{G}(s) ]\\
&=k_{a2D}'\left[1-\frac{\beta_1}{\lim_{s\to 0} [s+\beta-sF_1(s_1|s_1)z-\beta_2F(s_1|s_1;z)]} \right]\\
&=k_{a2D}'\left[1-\left(1+\frac{\beta_2}{\beta_1}\lim_{z\to 1}\frac{z^2/6}{P(s_0|s_0;z)-1}\right)^{-1} \right].\\
}
Finally, by taking the small $z$ expansion together with Eq. \eqref{Pz1}, we obtain the asymptotic rate coefficient expression:
\meq{DLks}{\lim_{s\to 0}s\hat{k}'_{2D}(s)&=\frac{2\pi k_{a2D}'}{P_a\sqrt{3}}\left\{\ln\left[\frac{12\exp\{2\pi(1/P_a-1)/\sqrt{3}\}}{1-z} \right]\right\}^{-1}\\
&=\frac{4\pi D}{\ln\left[E(1+\frac{4D}{l^2s}) \right]}\\
&\approx\frac{4\pi D}{\ln(C_l/s)}.}

By comparing the lattice and continuum rate coefficient, we found that the asymptotic expression in Eq. \eqref{DLks} for the diffusion-influenced scheme is the same as its continuum counterpart shown in Eq. \eqref{asympks}, while the time domain expression in Eq. \eqref{knl} for the activation-limited scheme is consistent with the continuum counterpart shown in Eq. \eqref{kck}. To match the lattice and continuum rates, we need to impose the equality $C_l=C_c$. It then implies that the reaction  probability should be chosen as
\eq{newP}{P_a=\left[1+\frac{\sqrt{3}}{2\pi}\left(\ln(f^2/12)+\frac{4\pi D}{k_{a2D}}-2\gamma\right) \right]^{-1},}
where $f=l/R$ denotes the ratio of voxel to the molecule size. 
Since probability $P_a$ is positive, it sets an additional constraint:
\meq{}{
\ln f+\frac{2\pi}{\kappa}&>-\frac{\pi}{\sqrt{3}}+\frac{\ln 12}{2}+\gamma=C_1
}
To satisfy the last inequality, $f=l/R$ has to be adapted according to the value of $\kappa$. Since $\kappa$ is always positive, we only need to set a lower bound expression for the voxel size:
\meq{ineq}{\ln f&>C_1-\frac{2\pi}{\kappa}>C_1,\\
f&>\exp(C_1),\\
l&>\exp(C_1)R\approx 1.005887R.\\
}

In 3D MLM, accurate reaction kinetics requires the voxel size of HCP lattice to be larger than the molecule by $l\approx 1.02086R$ \cite{chew2018}. If an HCP lattice volume compartment is bounded by a triangular lattice surface, the 3D voxel size condition would still satisfy Eq. \eqref{ineq}. Therefore, all surface and volume voxels in the model can adopt the same HCP voxel size. 

The accuracy of the lattice theory can be verified by comparing the theoretical values for the rate coefficient $k_{2D}'(t)$ with the simulated values. We obtained the theoretical rate coefficient from the numerical inverse Laplace transform of Eq. \eqref{kslaplace}. We simulated the reaction in Eq. \eqref{irrrxn} with Spatiocyte at both the activation-limited ($\kappa=0.01\cdot 4\pi$) and the diffusion-limited ($\kappa=100\cdot 4\pi$) regimes. We logged the number of surviving $A$ and used it to calculate the rate coefficient. The discretization of the time derivative in Eq. \eqref{STirrA} gives the formula for discrete rate coefficient:
\eq{ktcal}{k_{j+1}=-\frac{S_{j+2}-S_{j}}{[B]S_{j+1}\ (t_{j+2}-t_{j})},\text{ for }j\in \mathbb{Z}^+,} 
where $j$ is the index of the discretized $S_A$ and $t$. 
The boundary cases are computed as
\eq{}{k_1=-\frac{S_{2}-S_{1}}{[B]S_{1}\ (t_2-t_1)},\ k_N=-\frac{S_{N}-S_{N-1}}{[B]S_{N}\ (t_{N}-t_{N-1})},}
where $N$ denotes the final time step. 

\begin{figure}[!h] 
\centering
    \begin{subfigure}[t]{0.49\textwidth}
  	\centering
    \includegraphics[width=\textwidth]{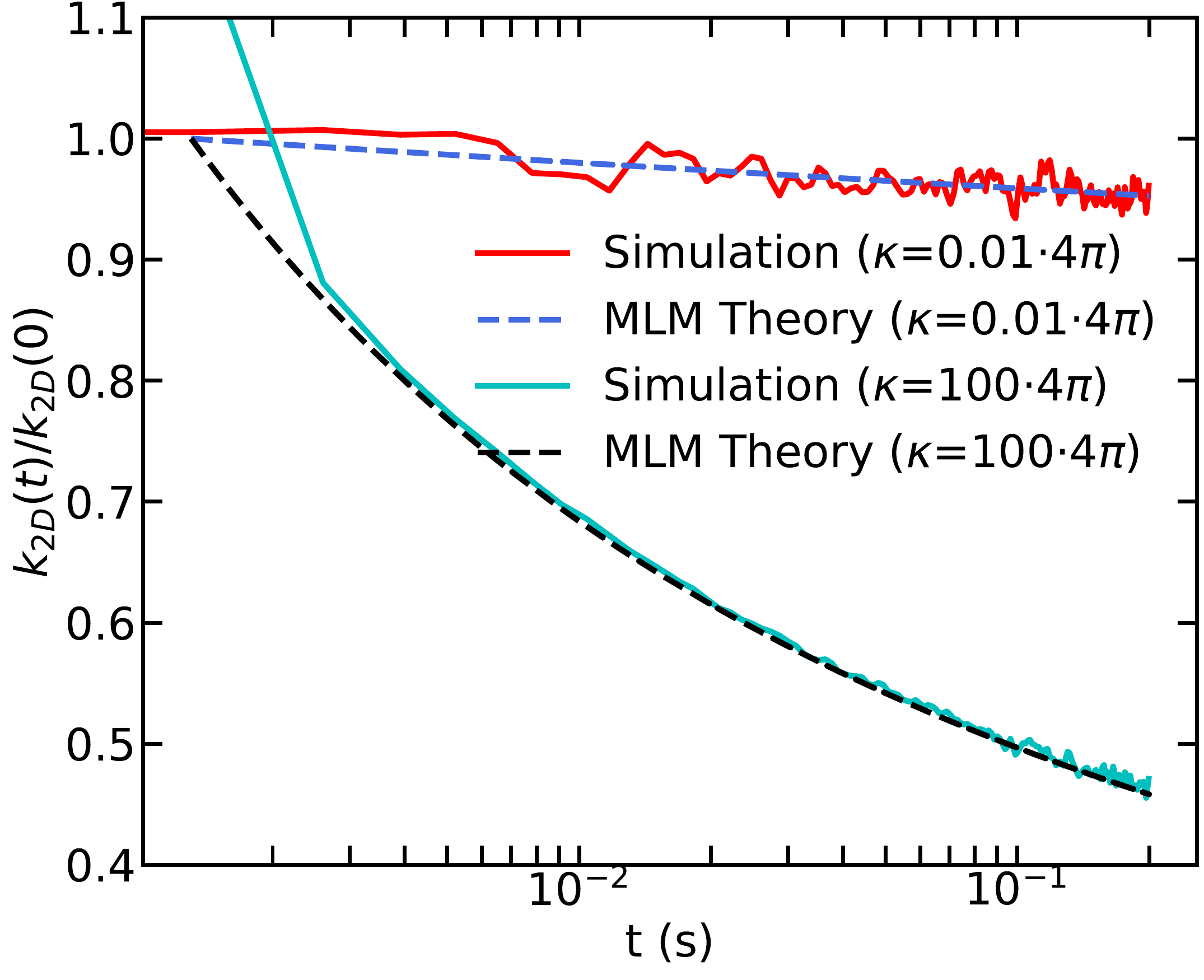}    
  	\caption{}
	\label{fig:kt1}
	\end{subfigure}
	\hfill
	\begin{subfigure}[t]{0.49\textwidth}
   	\centering	  
    \includegraphics[width=\textwidth]{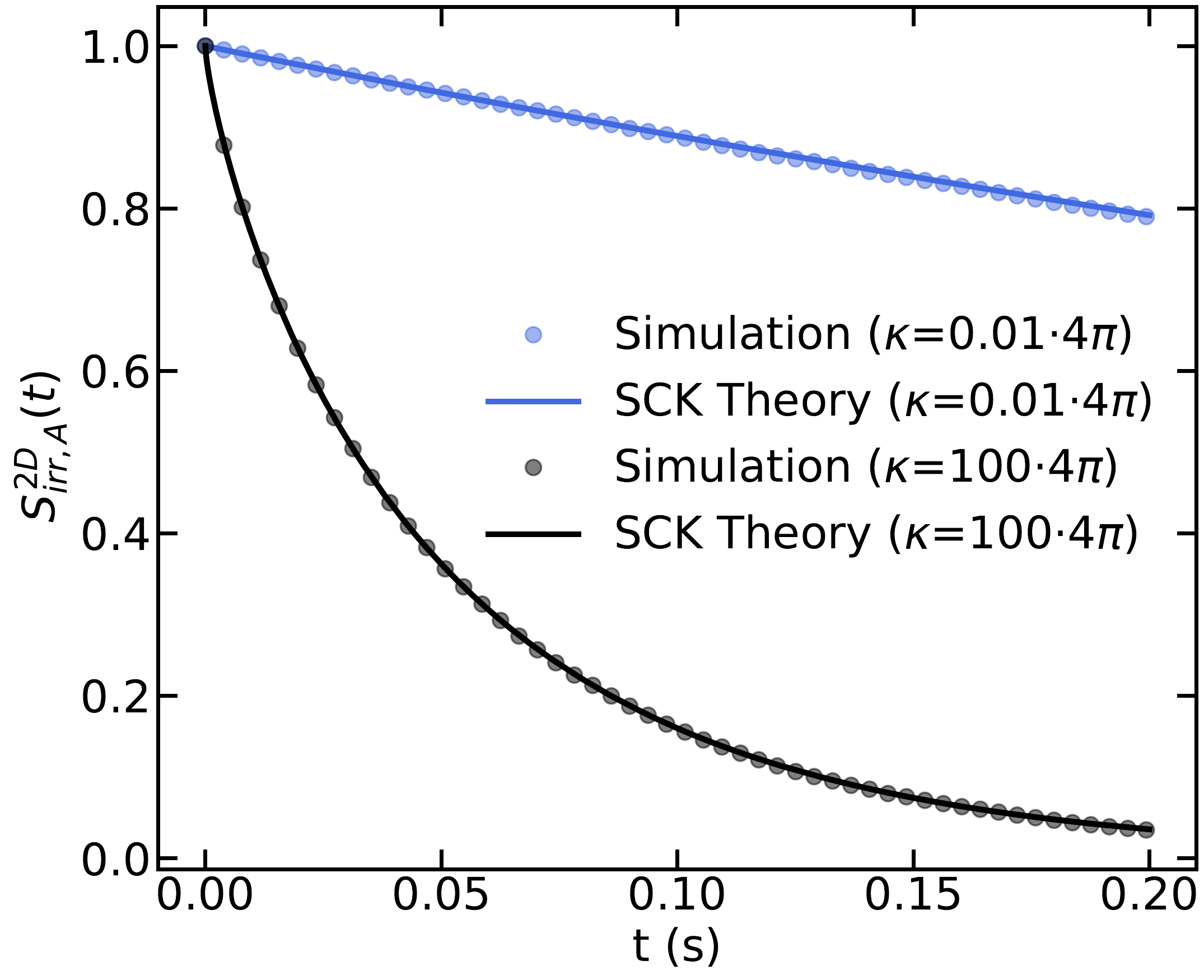}    
  	\caption{}
	\label{fig:st1}   	
	\end{subfigure}
	
  \caption{Comparison of on-lattice simulations with on- and off-lattice theories for surface-surface reaction $A+B \xrightarrow{} B$. (a) Simulated on-lattice time-dependent rate coefficients (solid lines) compared with on-lattice MLM theory in Eq. \eqref{kslaplace} (dashed lines). For better visualization of the time-dependent behavior of the two extreme cases, the simulated and theoretical lines are normalized by the initial theoretical value. (b) Simulated on-lattice survival probability of $A$ (points) compared with off-lattice SCK theory in Eq. \eqref{2DirrST} (solid lines). Activation-limited ($\kappa=0.01\cdot 4\pi$) and diffusion-limited ($\kappa=100\cdot 4\pi$) cases are indicated by the top and bottom lines respectively. Simulations were performed with Spatiocyte and the following parameters: Area = $(6.5\times 6.5)\ \mu \mathrm{m}^2$, $R=0.01\ \mu \mathrm{m}$, $l=0.01\times1.0209\ \mu \mathrm{m}$, $D_A=1$, $D_B=0\ \mu \mathrm{m^{2} s^{-1}}$, $N_a=N_b=423$, duration = 0.2 s, logging interval=$50t_d$}
\end{figure}

We compared the rate coefficients from the simulations with the theoretical values from Eq. \eqref{knl}. Figure \ref{fig:kt1} displays good agreements for both at activation-limited ($\kappa=0.01\cdot 4\pi$) and diffusion-limited ($\kappa=100\cdot 4\pi$) regimes for $t\gg t_d$. Next, we compared the simulated survival probability of the same reaction with the continuum-based theory, where the values are numerically evaluated according to 
\eq{2DirrST}{S_{irr,A}^{2D}(t,[B])=\exp\left[-[B]\int_0^t k_{2D}(\tau)d\tau \right].}
As shown in Figure \ref{fig:st1}, the simulated results overlap almost perfectly with the continuum-based theory, thus, confirming the accuracy of MLM.

\subsubsection{Reversible reaction}
\paragraph{}
Accurate simulation of reversible reaction $A+B \underset{k_{d2D}}{\stackrel{k_{a2D}}{\rightleftharpoons}} C$ according to the SCK model needs to satisfy the local detailed balance. This is achieved in MLM by adopting a rate constant $k_{d2D}'$ for the dissociation reaction such that the relation
\eq{}{\frac{k_{a2D}'}{k_{d2D}'}=\frac{k_{a2D}}{k_{d2D}},
}
is satisfied.

\begin{figure}[!h]   
\centering
  \includegraphics[width=.5\linewidth]{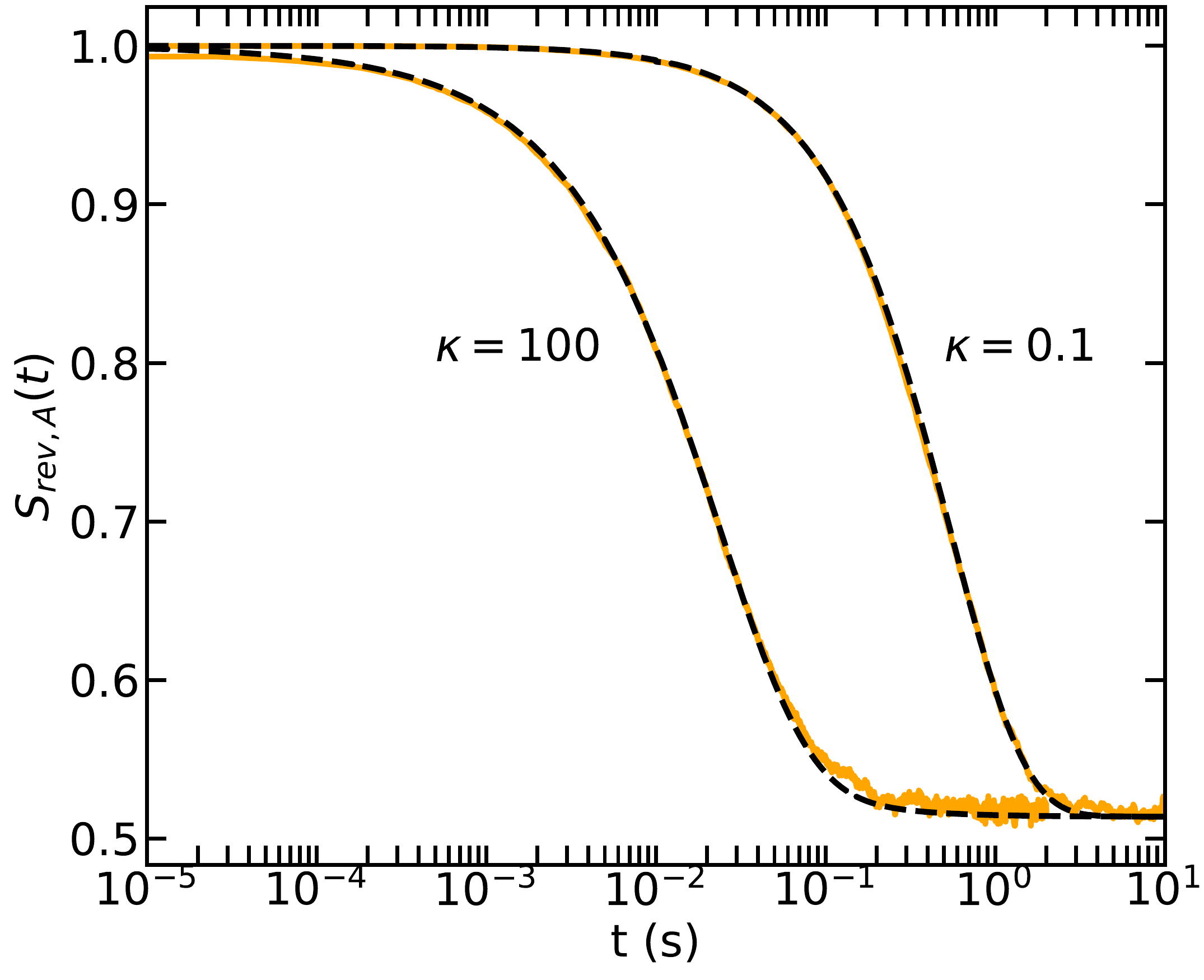}    
  \caption{ Survival probability of $A$ in the surface-surface reaction $A+B \underset{k_{d2D}}{\stackrel{k_{a2D}}{\rightleftharpoons}} C$. Dashed curves are the values calculated according to the MPK1 theory given in Eq.  \eqref{Stre}, solid lines are the simulation results of Spatiocyte. Association rates in the activation-limited ($\kappa=0.1$) and diffusion-limited ($\kappa=100$) regimes are chosen. Simulation parameters: $k_{d2D}=10k_{a2D}$, surface area = $(6.5\times 6.5)\ \mu \mathrm{m}^2$ with periodic boundary, $R=0.01\ \mu \mathrm{m}$, $l=0.01\times1.0209\ \mu \mathrm{m}$, $D_A=D_C=0\ \mu \mathrm{m^{2} s^{-1}}$, $D_B=1\ \mu \mathrm{m^{2} s^{-1}}$, $N_b=20,N_b=401$, duration =10 s.
  }
\label{streversible}
\end{figure}

We perform numerical simulations to confirm the ability of MLM to correctly reproduce the steady state and time-dependent behaviors in the reversible reaction. Association rates in the activation-limited ($\kappa=0.01$) and diffusion-limited ($\kappa=100$) cases were used in the simulation, while the dissociation rate $k_{d2D}$ is set to be 10 times larger than the association rate. Simulated result is compared with the MPK1 theory in Eq. \eqref{Stre}, obtained by numerical Laplace transform. The outcome shown in Figure \ref{streversible} indicates good agreement between the simulation and theory for time scales ranging from $t_d$ until equilibrium. 

\subsubsection{Generalization of MLM theory for other lattice arrangements}
\paragraph{}
The expression of MLM parameter derived for triangular lattice can be generalized to other lattice arrangements that adopt MLM. 
In general, the variable $C_l$ in Eq. \eqref{knl} takes the form of 
\eq{}{
C_l=4b_2D\exp\left\{(1/P_a-1)/b_1\right\}/l^2,}
where $b_1$ and $b_2$ are coefficients present in the highest order term of the generating function $P(s_0|s_0;z)$:
\eq{}{P(s_0|s_0;z)\approx b_1\ln\left(\frac{b_2}{1-z}\right).} 
On the other hand, the reaction  probability has the following general form:
\eq{}{P_a=\left[1+b_1\left(\ln(1/b_2)+\frac{4\pi D}{k_{a2D}}-2\gamma\right) \right]^{-1}.}
The expression for the probability has the following constraint on the voxel size:
\eq{}{l>\exp \left(\gamma-\frac{1}{2b_1}+\frac{\ln b_2}{2} \right)R.}

Here as an example, we consider the square lattice, a popular lattice choice to simulate surface reactions. The coefficients for square lattice are given as $b_1=1/\pi$ and $b_2=8$ (Eq.A.187 in \cite{hughes1995random}). 
The corresponding reaction  probability is
\eq{}{P_a=\left[1+\frac{1}{\pi}\left(\ln(1/8)+\frac{4\pi D}{k_{a2D}}-2\gamma\right) \right]^{-1},}
with the voxel size constrained by 
\eq{}{l>1.04722R.}
Therefore, to recapitulate the correct continuum rate, the voxel size in square lattice has to be about $5\%$ larger than the molecule size. This voxel size is substantially larger than the $0.6\%$ required by the triangular lattice. The different voxel size requirements reflect the influence of lattice arrangement on the first-passage time behavior and emphasize the importance of choosing the right MLM parameters to generate accurate reaction kinetics. 

\subsection{1D volume-surface adsorption}\label{ss2}
\paragraph{}
We next formulate the on-lattice 1D rate coefficient according to the SCK model and apply the rate expression to the problem of volume-surface adsorption. 

In 1D lattice, the generating function for the voxel occupancy probability from origin to origin is defined as \cite{montroll1965random}
\eq{}{P(s_0|s_0;z)=\frac{1}{\sqrt{1-z^2}}.}
\\The corresponding first passage time distribution, obtained by the relation $F(s_0|s_0;z)=1-1/P(s_0|s_0;z)$ (Eq. (I.18) in \cite{montroll1965random}), is given by 
\eq{GF1D}{F(s_0|s_0;z)=1-\sqrt{1-z^2}.}
\\Substituting Eq. \eqref{GF1D} into Eq. \eqref{ALGF} yields the generating function for the rebinding-time probability distribution
\eq{}{H(s_0|s_1;z)\approx \frac{-\sqrt{2(1-z)}}{P_a},}
where we consider only the highest order term $\sqrt{1-z}$ in the limit of $z\to 1$.

The corresponding large $n$ coefficient is obtained from the generating function according to the rule given in Figure VI.4 of \cite{flajolet2009analytic} as:
\eq{Hn1D}{H_n(s_0|s_1)\approx \frac{1}{P_a\sqrt{2\pi n^3}}.}
Applying Eq. \eqref{Hn1D} to the Noyes' rate formula in Eq. \eqref{ktlattice}, we obtain the asymptotic form for the 1D rate coefficient:
\eq{1dnt}{k_m^{1D}\approx k_{a1D}'\frac{\sqrt{2}}{P_a\sqrt{\pi m}.},
}
Using the definitions of initial lattice rate constant given in Eq. \eqref{lir1d} and the 1D simulation step size $ml^2=2Dt$, we have the rate expression as a function of time: 
\eq{asym1Dr}{k_{1D}'(t\to \infty)\approx 2\sqrt{\frac{D}{\pi t}}.}
Note that Eq. \eqref{asym1Dr} shares the same time-dependent form as the continuum-based theory given in Eq. \eqref{kt1dlt}. 

\begin{figure}[!t] 
\centering
    \begin{subfigure}[t]{0.49\textwidth}
  	\centering
    \includegraphics[width=\textwidth]{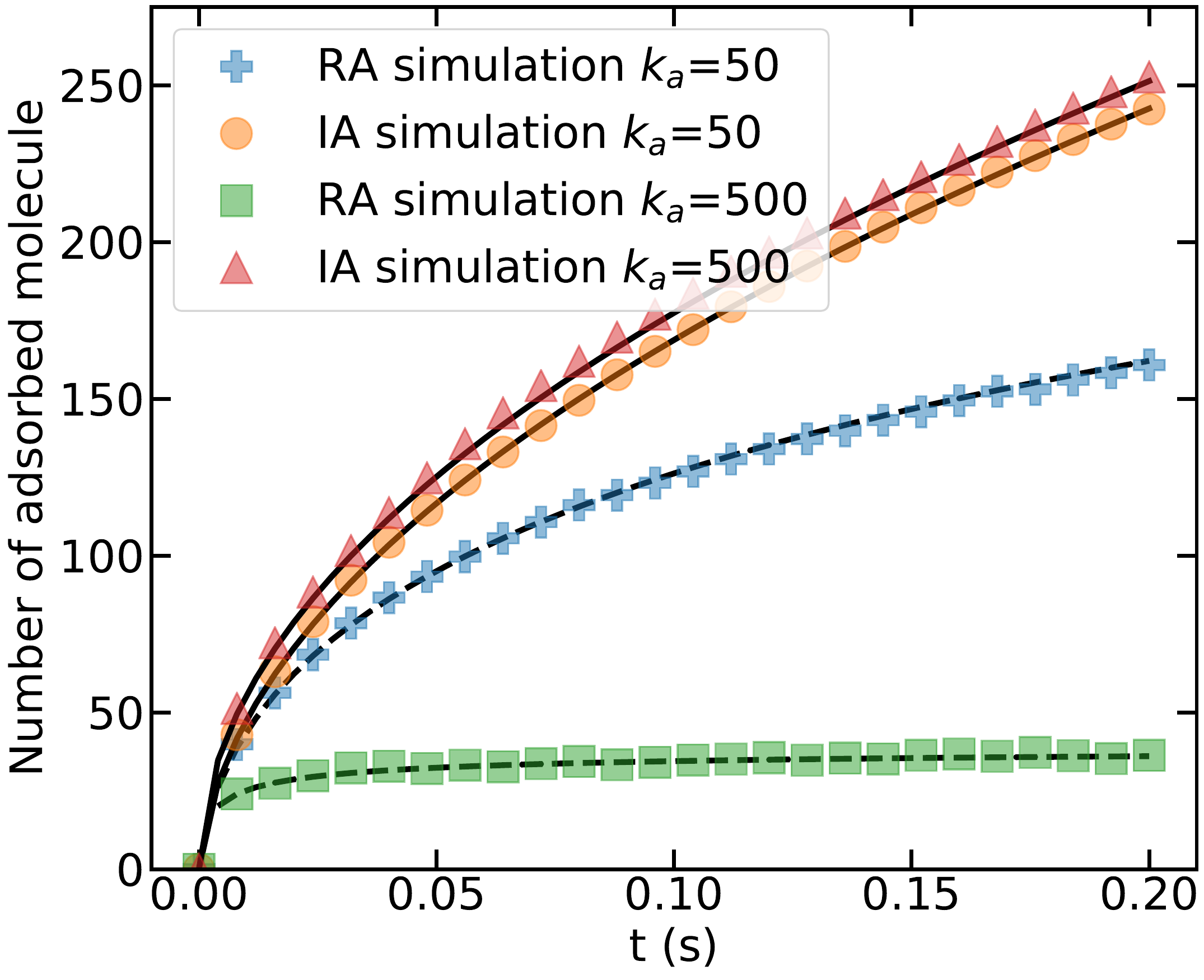}    
  	\caption{}
	\label{fig:ct}
	\end{subfigure}
	\hfill
	\begin{subfigure}[t]{0.49\textwidth}
   	\centering	  
    \includegraphics[width=\textwidth]{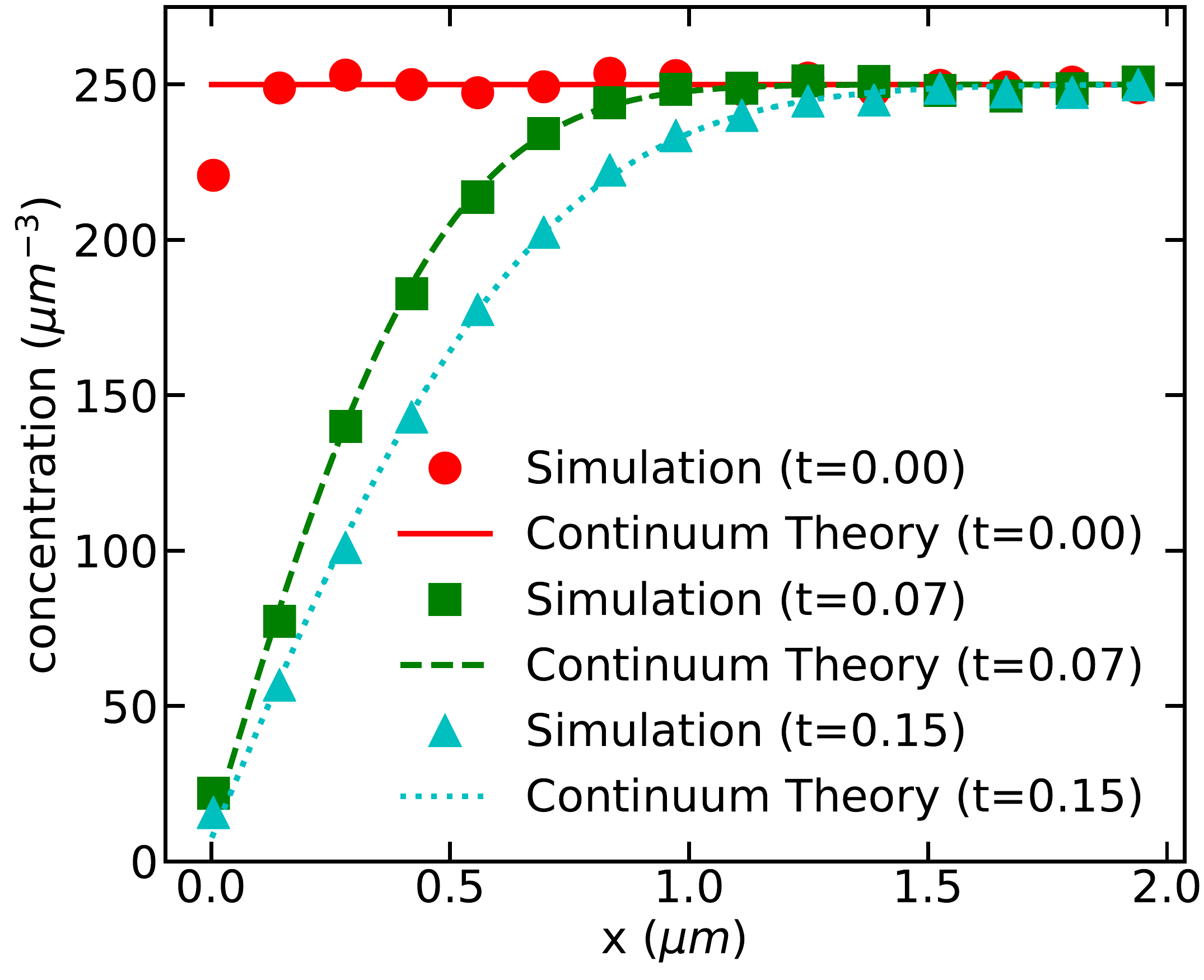}    
  	\caption{}
	\label{fig:cx}   	
	\end{subfigure}
	
	\caption{(a) Time series of adsorbed molecules simulated with irreversible (IA, triangle and circle markers) and reversible (RA, plus and square markers) adsorptions. In each case, strong ($k_{sa}=500\ \mu \mathrm{m}s^{-1}$) and weak ($k_{sa}=50\ \mu \mathrm{m}s^{-1}$) adsorption rates were tested. In the reversible adsorption, the membrane dissociation rates are $k_{sd}=62.5\text{ and }6250\ s^{-1}$, corresponding to the association rates $k_{sa}=50\text{ and }500\ \mu \mathrm{m}s^{-1}$, respectively. Solid and dashed lines represent the continuum-based values according to the irreversible and reversible reaction formulas in Eqs. \eqref{Nsa} and \eqref{Nsad}, respectively. (b) The concentration profile of cytosolic $A$ along the axis perpendicular to the adsorbing surface at $x=0$ for the given time points. The adsorption is irreversible with the rate $k_{sa}=50\ \mu \mathrm{m}s^{-1}$. Theoretical lines shown are according to the continuum-based theory in Eq. \eqref{gradient}. Simulation parameters: $l=0.01\ \mu \mathrm{m}$, $D_A=1\ \mu \mathrm{m^{2} s^{-1}}$, and initial number of cytosolic molecule  $N_a=1000$.}
\end{figure} 

For volume-surface adsorption, the definitions for initial adsorption rate constant in Eq. \eqref{lirsa} and the 3D simulation step size relation $nl^2=6Dt$ are used in Eq. \eqref{1dnt}. The resulting adsorption rate coefficient is given as
\eq{}{k_{sa}'(t\to \infty)\approx \frac{1}{2\sqrt{2}}\sqrt{\frac{D}{\pi t}},} 
which shares the same long-time scaling behavior with the continuum-based theory in Eq. \eqref{ktadlt} up to the same order. In contrast to the 3D and 2D cases, the long-time expression for the 1D rate coefficient does not depend on the reaction probability and the voxel size. 

Since the long-time rate coefficient has the same form in both lattice and continuous spaces, we only need to match the initial lattice rate constant $k_{sa}'$ with the adsorption rate constant $k_{sa}$ in continuum. This gives an expression for the reaction probability in terms of the adsorption rate constant, diffusion coefficient and voxel size (derivation is shown in Appendix \ref{lsa}):
\eq{}{P_a=\frac{\sqrt{2}k_{sa}l}{\sqrt{3}D},}

To examine the accuracy of MLM in simulating the adsorption kinetics, we performed Spatiocyte simulations using the derived expression for the reaction probability. We used a large number of cytosolic $A$ molecules in a cuboid compartment with a cross sectional area $(1\ \mu \mathrm{m})^2$ and length $4\ \mu \mathrm{m}$. An adsorbing plane is placed in the middle of the cuboid compartment, allowing adsorption from both sides of the surface. The number of adsorbed molecules at each time step is monitored. 

Figure \ref{fig:ct} shows the time series of $A$ on the adsorbing plane for irreversible (adsorption only) and reversible (adsorption and desorption) reactions. Simulated results agree well with the expected values according to the continuum theories for the irreversible reaction in Eq. \eqref{Nsa} and reversible reaction in Eq. \eqref{Nsad}. The good fit can be seen at both strongly ($k_{sa}=500\ \mu \mathrm{m}s^{-1}$) and weakly ($k_{sa}=50\ \mu \mathrm{m}s^{-1}$) adsorbing rates. To examine the spatialtemporal concentration profile, we counted the number of cytosolic molecules near the adsorbing plane in the irreversible adsorption. The resulting concentration profile along the axis perpendicular to the adsorbing plane are shown in Figure. \ref{fig:cx}. The simulation results coincide very well with the curves of continuum-based theory in Eq. \eqref{gradient}. 

\section{Application of surface reactions} \label{s3}
\paragraph{}
A cytosolic molecule can react with a membrane-bound reactant via two possible pathways: it can either perform 3D diffusion in the cytoplasm and then directly react with the membrane-bound reactant exposed to the cytosol or it can bind first to the membrane and then perform 2D diffusion before reacting with the reactant. Both of these pathways are often adopted simultaneously in the cell. Previous works have investigated how each pathway contributes to the overall process \cite{adam1968reduction,berg1977physics,axelrod1994reduction,kholodenko2000cytoplasmic}. Here we apply the Spatiocyte scheme with the derived MLM expressions to simulate surface reactions comprising all dimensions. We study the contribution of each pathway to the overall reaction rate under the influence of different diffusivity and reactivity.

We consider a cuboid compartment of dimension $H\times L\times L$, depicting the cytoplasmic volume. The top surface of the cuboid is reflective, whereas the bottom surface represents an absorbing lipid membrane. Each of these surfaces has the area $L\times L$. 
Within the system, there are two elementary species, $A$ and $B$, with radius $r=0.005\ \mu m$. $A_c$ denotes the cytosolic state of $A$ that diffuses freely in the bulk at a rate of $D_c$. $A_c$ can reversibly associate with the membrane to become $A_m$: 
\eq{dm1}{A_c \underset{k_{sd}}{\stackrel{k_{sa}}{\rightleftharpoons}} A_m.}
The ratio of the membrane association constant over the dissociation constant is the equilibrium constant, $k_{sa}/k_{sd}=K_{eq}$. Upon the adsorption onto the membrane, $A_m$ performs 2D diffusion at a rate of $D_m$.
On the membrane, $B$ molecules are initialized to be immobile and randomly distributed with concentration $[B]_0$.

$A$ can react with $B$ via the 3D pathway:
\eq{dm3}{A_c+B \xrightarrow{k_{a3D}} AB,}
or the 2D pathway:
\eq{dm2}{A_m+B \underset{k_r}{\stackrel{k_{a2D}}{\rightleftharpoons}} AB.}
$k_{a\{2D,3D\}}$ denotes the intrinsic association rate constants for 2D and 3D reactions, whereas $k_r$ represents the dissociation rate constant. 

To quantify the dominance of the 2D pathway, we measured the fraction of the 2D equilibrium rate in the total reaction rate, as in \cite{axelrod1994reduction}: 
\eq{f2d}{f_{2D}=\frac{k_{on2D}}{k_{on2D}+k_{on3D}}=\frac{1}{1+k_{on3D}/k_{on2D}}.}
$k_{on\{2D,3D\}}$ represents the macroscopic effective rates for the 2D and 3D association reactions. The  $k_{on3D}/k_{on2D}$ ratio is calculated using the simulated equilibrium concentrations according to the formula 
\eq{}{\frac{k_{on3D}}{k_{on2D}}=\frac{1}{[A_c]_{eq}}\left(\frac{k_r[AB]_{eq}}{k_{a2D}[B]_{eq}}-[A_m]_{eq}\right),} which is derived by solving the rate equations for Eqs. \eqref{dm2} and \eqref{dm3} at equilibrium. 

We examined the dominance of the 2D pathway with changes in $D_c/D_m$, $[B]_0$, and the association reaction probability, $P_{a\{2D,3D\}}$ for the 2D and 3D pathways. We fixed other variables such as the sizes of the system and molecule, $k_r$, $K_{eq}$, $k_{a2D}/k_r$, and the initial concentration [$A_c$]. We used the typical cytosolic rate for $D_c$ ($10\ \mu m^2s^{-1}$) with $D_c/D_m$ ratio ranging from 1 to 1000. $K_{eq}=0.15\ \mu m$ and $k_{sd}=10\ s^{-1}$ are within the biologically realistic values \cite{Fulbright1993,Tolentino2008}. 

\begin{figure}[!t]   
\centering
  \includegraphics[width=.5\linewidth]{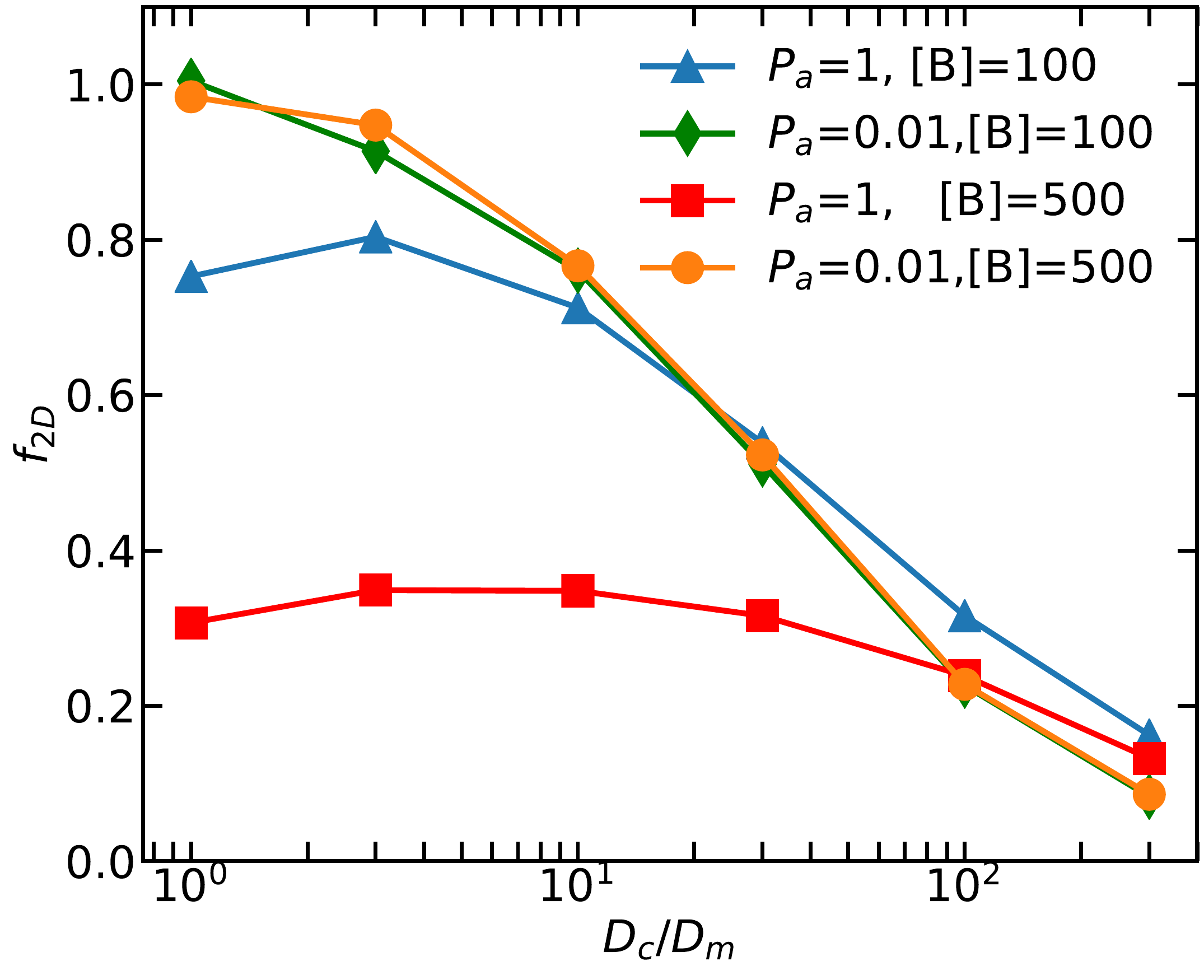}    
  \caption{Contribution of 2D reaction pathway in surface reactions. The fraction of 2D reaction pathway that contributes to the overall surface reaction is indicated by $f_{2D}$ and is plotted against $D_c/D_m$. The fraction is obtained at varying reaction probabilities, $P_a=P_{a2D}=P_{a3D}$ and concentration of the membrane-associated reactant [B] (unit $\mu m^{-2}$). Simulation parameters: $R=0.01\ \mu m$, $l=0.01\times1.0209\ \mu \mathrm{m}$, $L=1\ \mu m$, $H=2L$, $D_c=10\ \mu \mathrm{m^{2} s^{-1}}$, $[A_c]=5 \mu M=3000 \mu m^{-3}$, $K_{eq}=0.15\mu m$, $k_{sd}=10\ s^{-1}$, $k_{a2D}/k_r= 0.001\ \mu m^2$.   
  }
\label{demo}
\end{figure}

From the simulation results in Figure \ref{demo}, we can observe the overall decreasing trend of $f_{2D}$ as the ratio $D_c/D_m$ increases. The exact value of $f_{2D}$ depends on the reaction probability and the concentration of reactant, $[B]_0$. When the association reaction is diffusion limited ($P_{a2D}=P_{a3D}=1$) and the reactant concentration is low ($[B]_0$=100 $\mu m^{-2}$), $f_{2D}$ becomes more than $50\%$ for $D_c/D_m$ between 1 and 30. When $D_c/D_m > 30$, the 3D pathway becomes dominant instead. At very high $[B]_0$ ($500\ \mu m^{-2}$), the 3D pathway is dominant for all ratios of $D_c/D_m$. When the association reaction is activation-limited ($P_{a2D}=P_{a3D}=0.01$), $f_{2D}$ is still larger than $50\%$ for $D_c/D_m$ in the range [1,30], and becomes less than $50\%$ when the ratio is higher than 30. However, unlike in the diffusion-limited case, $f_{2D}$ in activation-limited reaction is less sensitive to the changes in $[B]_0$.

In typical cells, membrane-associated molecules diffuse 10 to 100 times slower than their cytosolic counterpart. In such a condition, our simulation results imply the following: the 2D reaction pathway will dominate the overall reaction, provided the concentration of membrane-associated reactant is low, its diffusion on the membrane is fast and the reaction is activation-limited. Conversely, the 3D reaction pathway will become dominant when the diffusion of membrane species is slow or when the membrane-associated reactant is abundant and reacts with high probability upon collision. 

\section{Discussion and conclusion} \label{s4}
\paragraph{}
 MLM surface reactions have not been verified in terms of their consistency with continuum-based theory. To address this issue, we used the theoretical framework of MLM \cite{chew2018} to derive the correct expressions for 2D surface-surface reaction and 1D volume-surface adsorption on lattice. By employing the SCK model and the random walk theory, we showed that the 2D lattice reaction exhibits the same long-time behavior as the continuum-based theory. After equating the on-lattice rate expression with that of the continuum theory, we obtained the formula for the reaction probability in terms of physical and lattice parameters. 
 
 Furthermore, the positively valued reaction probability imposes an additional constraint on the voxel size: it should be larger than the molecule at least by about $0.6\%$ for the triangular lattice and by $5\%$ for the square lattice. These constraints also meet the minimum voxel size requirement of the corresponding lattice arrangement in 3D \cite{chew2018}. If the voxel size is exactly the same as the molecule, the simulated time-dependent reaction kinetics will deviate from the expected behavior in continuum. Such deviations should be carefully considered especially when simulating reactions containing nonlinear terms.
 
In 2D reversible reaction, we showed that correct equilibrium and time-dependent behaviors can be achieved by dissociating the substrate into an in-contact pair of product molecules, with a rate constant satisfying the local detailed balance. In 1D volume-surface adsorption, the long-time asymptotic behavior of MLM has the same form as in the continuum-based theory. The Spatiocyte scheme also generated spatiotemporal adsorption kinetics that is consistent with continuum theory when the correct expression for reaction probability was used. 

Finally, we studied the contribution of a 2D reaction pathway in a surface reaction model with Spatiocyte simulations. We found that the dominant surface reaction pathway can be sensitive to the surface reactant concentration, intrinsic reaction rate and the relative diffusivity of reactants between the bulk and the surface. For example, the 2D reaction pathway would play a significant role in regulating the overall rate for a system that has a sparse membrane-associated reactant with activation-limited rate constants.

The main advantage of MLM when modeling intracellular reaction-diffusion processes is its ability to capture the microscopic properties of molecules directly without incurring high computational cost. As an illustration, it only takes minutes for Spatiocyte to simulate thousands of molecules with a time step of $\mu$s for a duration of seconds on a single CPU core (see performance in \cite{chew2018}). Spatiocyte takes physical quantities comprising molecule size, diffusion coefficient and intrinsic reaction rate as input, and generates time-series output such as molecule copy number and trajectory.

At present, Spatiocyte supports surface reactions with various geometries at the cellular scale. It has been successfully used to capture the influence of microscopic effects on the behavior of cells at the macroscopic scale. These include the formation of a high density ring over the entire bacterium cell membrane as a result of transient membrane association and rebinding of proteins \cite{Arjunan2010}, the clustering of proteins on the red blood cell membrane from oxidative stress \cite{Shimo2015} and the oligomerization of receptors and its influence on ligand binding kinetics \cite{watabe2018simulation}. As the spatiotemporal resolution of imaging techniques continue to advance \cite{hell2015}, time-dependent reaction kinetics and molecular trajectories will become more accessible. These high resolution experimental data coupled with efficient microscopic simulation techniques such as MLM will provide a complementary way to investigate mechanisms underlying various biological reaction-diffusion processes. 

The uniform voxel size adopted by MLM reduces computational complexity and consequently, contributes to its low computational cost. However, in surface systems requiring realistic simulation of distinct-sized molecules with non-spherical structures, additional considerations would be needed for MLM to be applied. One potential solution is to reduce the voxel size and let a single molecule occupy more than one voxel according to its size and shape. Alternatively, we can represent molecules with distinct shapes and sizes off-lattice and perform hybridized simulation with on-lattice molecules. The implementation and accuracy of such schemes compared to fully off-lattice methods would require further examination. Another future milestone for MLM is to establish and verify its consistency in highly crowded environments. The on-lattice rate has to be reformulated to account for the many-particle interaction. The resulting lattice theory should then be compared and matched with the continuum-based theory.

\section*{Acknowledgements}
\paragraph{}
We thank K. Nishida for technical advice and support, and S. Kato for technical support and E-Cell 4 feature development. W.-X.C. acknowledges RIKEN for financially supporting his doctoral research as an International Program Associate. Part of this work was supported by JSPS KAKENHI Challenging Research (Pioneering) Grant No. 18H05371 to S.N.V.A.  
 
W.-X.C., K.K., K.T. and S.N.V.A. designed research; W.-X.C. performed research;  W.-X.C., K.K., M.W., S.V.M., and S.N.V.A. analyzed data; and W.-X.C. and S.N.V.A wrote the manuscript. All authors read and commented on the manuscript.

\appendixtitleon
\numberwithin{equation}{section}
\begin{appendices}
\section{Lattice initial rate constant}\label{LIR}
\paragraph{}
In this section, we provide the derivation of the lattice initial rate constant for heterodimerization and homodimerization reaction. Solution for general regular lattice arrangement, triangular lattice and square lattice is shown. Consider two reacting species $A$ and $B$, in which $A$ are stationary and $B$ are diffusing with relative diffusion coefficient $D$. The two species associate irreversibly to form a complex with initial rate constant $k_a'$: $A+B \xrightarrow{k_a'} C$. The number of reactions occurred in simulation time interval $t'$ in the continuum approximated by the law of mass action is related to the lattice space by:
\eq{}{\Delta [C]=k_a'[A][B]t'=\text{number of reaction on lattice as a function of reaction probability},}
where $[\ ]$ denotes the molecule concentration.
\subsection{Initial rate constant for 2D reaction}
\paragraph{}
On 2D lattice, the number of reactions occurred in interval $t'$ according to the continuum-based framework is given as 
\eq{}{\Delta N_C=\frac{k_{a2D}'N_AN_Bt'}{S},}
where $N_i$ is the number of molecules of species $i$, $\Delta N_C$ is the changes in molecule number $N_c$ and $S$ is the surface area. 

Whereas the number of reaction in a step interval $t'$ on lattice can be estimated as 
\eq{}{\Delta N_C=\frac{P_a'N_BN_A}{N_{sv}},}
where $N_{sv}=Sd/(\pi l^2/4)$ is the number of surface voxels and $d$ is the packing density of the lattice type, and $P_a'=P_a\alpha$ is the actual reaction  probability during the encounter.

In the activation-limited scheme (see section II.B. in \cite{chew2018} for detail description of the scheme), we have $t'=t_d$ and $P_a'=P_a$. 
Thus, we have
\meq{generalrate}{k_{a2D}'&=\frac{P_a S}{t_d N_{sv}},\\
&=\frac{\pi P_a D}{d}.}
Note that the physical unit of $k_{a2D}'$ is [$\mathrm{L^2T^{-1}}$], consistent with the continuum rate constant.
\\For triangular lattice we have $d=\pi \sqrt{3}/6$ and $N_{sv}=2S/(\sqrt{3}l^2)$. Thus the lattice initial rate is given as
\eq{lir2d}{k_{a2D}'=2\sqrt{3}P_aD,}
valid for both the activation-limited and diffusion-influenced schemes.
\\As for square lattice, we have $d=\pi/6$, and the initial rate is 
\eq{slka}{k_{a2D}'=6P_aD.}

In homodimerization reaction $A+A \xrightarrow{k_a'} C$, the number of reactions according to continuum framework is 
\eq{}{\Delta N_C=\frac{k_{a2D}'N_A(N_A-1)t'}{S}.}
Whereas on lattice we have
\eq{}{\Delta N_C=\frac{P_a'N_A(N_A-1)}{2N_{sv}}.}
From these two equations, the lattice rate constant is derived as 
\meq{}{k_{a2D}'&=\frac{P_a S}{2t_d N_{sv}},\\
&=\frac{\pi P_a D}{2d},}
which differs from Eq. \eqref{generalrate} by a factor of 2.
As for the triangular lattice, the lattice rate constant is given as
\eq{}{k_{a2D}'=\sqrt{3}P_aD,}
in which the relative diffusion coefficient $D$ is defined as the sum of the two diffusion coefficients $D_A$.

\subsection{Initial rate constant for 1D reaction}
\paragraph{}
The number of reactions in interval $t'$ according to the continuum framework is given as 
\eq{}{\Delta N_C=\frac{k_{a1D}'N_AN_Bt'}{L},}
where $L$ denotes the length of the 1D system. 

To be compatible with the continuum framework, we have the following assumptions in the derivation of the lattice rate constant: (i) each voxel can accommodate more than one molecule; (ii) molecules A are static whereas molecules B are mobile with relative diffusion coefficient D. 
Then the number of reactions happens in a simulation step on lattice can be approximated by 
\meq{}{\Delta N_C=\frac{P_aN_AN_B}{N_L},}
where $N_L=L/l$ denotes the number of lattice voxels in length $L$.

Finally, the initial lattice rate constant is given by 
\meq{lir1d}{k_{a1D}'&=\frac{P_a l}{t_d},\\
&=\frac{2DP_a}{l},}
with unit of [$\mathrm{LT^{-1}}$].

\subsection{Initial rate constant for volume-surface adsorption}\label{lsa}
\paragraph{}
Consider a cuboid compartment with an adsorbing plane in the middle. Molecules $A$ diffuse in the bulk with diffusion coefficient $D_A$. Adsorption occurs on both sides of the plane. 
\\According to the continuum theory, the number of adsorbed molecules in time step $t'$ is approximated by
\eq{}{\Delta N_s=\frac{2k_{sa}'N_At'S}{V},}
where $N_s$ is the number of molecules adsorbed, $N_A$ is the initial number of molecule in the bulk, $S$ is the area of the plane and $V$ is the volume of the cuboid compartment.

In the case of HCP lattice arrangement, the number of adsorption to the plane is approximated by
\meq{chg2}{\Delta N_s &= P_a\frac{2N_{sv}}{N_v}\frac{3}{12}N_A,}
where $N_{sv}$ is the number of surface voxel (triangular lattice), $N_v=\sqrt{2}V/l^3$ is the number of volume voxel (HCP lattice) and $P_a$ is the reaction  probability. Note that 3/12 is the probability that a molecule adjacent to the plane hops to the plane in one step and $2N_s/N_v$ is the probability of a randomly distributed molecule $A$ adjacent to the plane. 

By equating these two expressions and solve for $k_{sa}$, we obtain
\eq{}{k_{sa}'=\frac{P_al}{2\sqrt{6}t_d}.}
Finally, with the diffusion time step definition $t_d=l^2/6D_A$, the initial lattice adsorption rate constant is expressed as
\eq{lirsa}{k_{sa}'=\frac{\sqrt{3}P_aD_A}{\sqrt{2}l},}
where the unit is [$\mathrm{LT^{-1}}$].

\section{Generating function derivation}
\subsection{2D rebinding-time probability distribution function}\label{ALHZ}
\paragraph{}
First, we express the generating function $H(s_0|s_1;z)$ as given in the main text into the following form
\meq{}{H(s_0|s_1;z)&=\frac{P_aF(s_0|s_0;z)}{z+F(s_0|s_0;z)(P_a-1)}\\
&=\frac{P_aF(s_0|s_0;z)}{z[1-F(s_0|s_0;z)(1-P_a)/z]}\\
&=\frac{P_a}{z}\left\{F(s_0|s_0;z)+\frac{(1-P_a)}{z}F(s_0|s_0;z)^2+\left[\frac{(1-P_a)}{z}\right]^2F(s_0|s_0;z)^3+...\right\}.}
Let $F(s_0|s_0;z)=1-1/P(s_0|s_0;z)$ as $1-x$ and $q=1-P_a$, we have
\meq{}{H(s_0|s_1;z)&=\frac{P_a}{z}\left\{1-x+\frac{q}{z}(1-x)^2+\left[\frac{q}{z}\right]^2(1-x)^3+...\right\},
}
in which regular $z$ terms are neglected since $z=1$. 
Finally, by rearranging the generating function in terms of $x$, we obtain
\meq{}{H(s_0|s_1;z)
&=P_a\left\{1+q\left[1+q+q^2+... \right]-x\left[1+2q+3q^2+... \right]+x^2q\left[1+3q+6q^2+... \right]+... \right\}\\
&=P_a\left\{1+\frac{q}{1-q}-x\sum_{n=1}^\infty nq^{n-1}+x^2q\sum_{n=1}^\infty \frac{n(n+1)}{2!}q^{n-1}+... \right\}\\
&=P_a\left\{\frac{1}{1-q}-\frac{x}{(q-1)^2}-\frac{x^2q}{(q-1)^3}-\frac{x^3q^2}{(q-1)^4}-... \right\}\\
&=P_a\left\{\frac{1}{1-q}-\frac{x}{(q-1)^2}\frac{1}{1-\frac{xq}{q-1}}\right\}\\
&=1-\frac{x}{P_a\left[1+\frac{x(1-P_a)}{P_a}\right]}.
}
\subsection{Voxel occupancy probability on triangular lattice}\label{trilat}
\paragraph{}
The voxel occupancy probability from origin to origin, $P_n(s_0|s_0)$ for the triangular lattice is given as \cite{guttmann2010lattice,hughes1995random}:
\eq{}{P_n(s_0|s_0)=\frac{1}{6^n}\sum_{j=0}^n\binom{n}{j}(-3)^{n-j}b_j,}
where
\eq{}{b_j=\sum_{k=0}^j\binom{j}{k}^2\binom{2k}{k}.}
The corresponding probability generating function is expressed as
\eq{Pz}{P(s_0|s_0;z)=\frac{6}{\pi z\sqrt{c}} \boldsymbol{\mathrm{K}}(k')}
where $c=(a+1)(b-1)$,
\eq{}{
a=\frac{3}{z}+1-\sqrt{3+\frac{6}{z}} \qquad \text{and} \qquad b=\frac{3}{z}+1+\sqrt{3+\frac{6}{z}}\ ,\\
}
\\and $\boldsymbol{\mathrm{K}}(k')$ is the complete elliptic integral of the first kind with 
\eq{}{
k'=\sqrt{\frac{2(b-a)}{c}}\ .
}
The asymptotic expansion of $P(s_0|s_0;z)$ in terms of the asymptotic form for $\boldsymbol{\mathrm{K}}(z)$ is derived as (see Eq.A.198 in \cite{hughes1995random}):
\eq{}{P(s_0|s_0;z)\approx \frac{\sqrt{3}}{2\pi}\ln[12(1-z)^{-1}]\{1+O(1-z)\}.}

\end{appendices}

\end{document}